\documentclass[conference]{IEEEtran}
\IEEEoverridecommandlockouts
\usepackage{cite}
\usepackage{amsmath,amssymb,amsfonts}
\usepackage{algorithmic}
\usepackage{graphicx}
\usepackage{textcomp}
\usepackage{xcolor}
\usepackage{bm,soul}
\usepackage{comment}
\usepackage{tikz}
\usepackage{pgfplots}
\usepackage{float}
\usepackage{color}
\usepackage{soul}
\usepackage{mathtools}
\usepackage{setspace}
\usepackage{cite}
\usepackage{scalerel}
\usepackage{comment}
\usepackage{mathtools}
\usepackage{subfigure}
\usepackage[acronym,shortcuts]{glossaries}
\def\BibTeX{{\rm B\kern-.05em{\sc i\kern-.025em b}\kern-.08em
    T\kern-.1667em\lower.7ex\hbox{E}\kern-.125emX}}
\def\overlinebold#1{\ThisStyle{\ooalign{%
  $\SavedStyle\mkern3mu\overline{\phantom{\mathrm{#1}}}$\cr $\SavedStyle\bm #1$}}}

\newacronym{$k$-NN}{$k$-NN}{k-Nearest Neighbors}
\newacronym{LSTM}{LSTM}{long short-term memory}
\newacronym{LOS}{LOS}{line of sight}
\newacronym{CSI}{CSI}{channel state information}
\newacronym[plural=RNNs,firstplural= recurrent neural networks (RNNs)]{RNN}{RNN}{recurrent neural network}
\newacronym[plural=CFRs,firstplural= channel frequency responses (CFRs)]{CFR}{CFR}{channel frequency response}
\newacronym[plural=CIRs,firstplural= channel impulse responses (CIRs)]{CIR}{CIR}{channel impulse response}
\newacronym{PLA}{PLA}{physical layer authentication}
\newacronym[plural=AoAs,firstplural= angles of arrival (AoAs)]{AoA}{AoA}{angle of arrival}
\newacronym[plural=AoAs,firstplural= angles of departure (AoDs)]{AoD}{AoD}{angle of departure}
\newacronym{MIMO}{MIMO}{multiple-input multiple-output}
\newacronym{SNR}{SNR}{signal-to-noise ratio}
\newacronym{ULA}{ULA}{uniform linear array}
\newacronym{UE}{UE}{user equipment}
\newacronym{BS}{BS}{base station}
\newacronym{RIS}{RIS}{reconfigurable intelligent surface}
\newacronym{ML}{ML}{machine learning}
\newacronym{mmwave}{mmWave}{millimeter-wave}
\newacronym{FA}{FA}{false alarm}
\newacronym{LRT}{LRT}{likelihood ratio test} 
\newacronym{LT}{LT}{likelihood test}
\newacronym{MD}{MD}{misdetection}
\newacronym{MAP}{MAP}{maximum posteriori probability}
\newacronym{awgn}{AWGN}{additive white Gaussian noise}
\newacronym{miso}{MISO}{multiple-input single-output}
\newacronym{ofdm}{OFDM}{orthogonal frequency division multiplexing}
\newacronym{ue}{UE}{user equipment}
\newacronym{parafac}{PARAFAC}{parallel factor}
\newacronym{cearc}{CEARC}{channel estimation with adaptive RIS configuration}
\newacronym{mbce}{MBCE}{MUSIC-based channel estimation}
\newacronym{mrc}{MRC}{maximum ratio combining}
\newacronym{mse}{MSE}{mean squared error}
\newacronym{CDF}{CDF}{cumulative distribution function}
\newacronym[plural=AoAs,firstplural=angles of arrival (AoAs)]{aoa}{AoA}{angle-of-arrival}
\newacronym{simo}{SIMO}{single-input multiple-output}
\newacronym[plural=AoDs,firstplural=angles of departure (AoDs)]{aod}{AoD}{angle-of-departure}
\newacronym{lmmse}{LMMSE}{linear minimum mean square error}
\newacronym{ls}{LS}{least-square}
\newacronym{zf}{ZF}{zero-forcing}
\newacronym{ce}{CE}{channel estimation}
\newacronym{dft}{DFT}{discrete Fourier transform}
\newacronym{music}{MUSIC}{multiple signal classification}
\newacronym{omp}{OMP}{orthogonal matching pursuit}
\newacronym{snr}{SNR}{signal-to-noise ratio}
\newacronym{pdf}{pdf}{probability density function}

\DeclareMathOperator*{\argmin}{arg\,min}

\begin{document}

\title{Security Analysis of RIS-Assisted Physical-Layer Authentication Over Multipath Channels
\thanks{M. Baldi is supported by the project SERICS (PE00000014) under the MUR National Recovery and Resilience Plan, funded by the European Union - Next Generation EU. M. Baldi and S. Tomasin are in part supported by European Union (EU) COST Action CA22168—Physical Layer Security for Trustworthy and Resilient 6G Systems (6G-PHYSEC). The work of L. Senigagliesi and S. Tomasin are supported by the European Commission through the Horizon Europe/Smart Networks and Services Joint Undertaking (JU SNS) Project ROBUST-6G under Grant 101139068.}
}

\author{Linda Senigagliesi$^{1}$, Anna V. Guglielmi$^{2}$, Marco Baldi$^{3}$, and Stefano Tomasin$^{2}$ \\
\IEEEauthorblockA{$^{1}$ETIS UMR 8051, CYU, ENSEA, CNRS, Cergy, France,
  $^{2}$University of Padova, Italy,\\
 $^{3}$Università Politecnica delle Marche, Ancona, Italy\\
	 email: linda.senigagliesi@ensea.fr,\{annavaleria.guglielmi, stefano.tomasin\}@unipd.it, m.baldi@staff.univpm.it
       }}

\maketitle

\begin{abstract}
In physical layer authentication, verification of a user's identity is based on the characteristics of the transmission channel through which signals are delivered to the authenticator (Bob). In this paper, we assume that the signals received by Bob pass through a \ac{RIS} (controlled by Bob) and that the legitimate transmitter (Alice) is equipped with one antenna. Conversely, the attacker (Trudy) has multiple antennas and uses precoding to deceive Bob's verification. Assuming that Trudy knows all the channel matrices, we first derive her optimal attack strategy. Then, we analyse the conditions under which the channel estimated by Bob is indistinguishable when either Alice or Trudy is transmitting. When Trudy has a single antenna, we show that the indistinguishability condition cannot be met when the channels to the RIS are the result of propagation over multiple paths. For single-path line-of-sight (LOS) conditions, instead, Trudy can impersonate Alice although transmitting from a different position. We verify these results numerically and assess the security of the considered scenario, even when the indistinguishability conditions cannot be met.
\end{abstract}


\section{Introduction}

Authentication is the process by which a receiver can verify the identity of a transmitter. Authentication mechanisms based on cryptographic algorithms remain secure provided that no computational breakthrough occurs, i.e., for new attack algorithms or the introduction of quantum computing. They typically entail high complexity, unsuitable in scenarios with limited power and computational resources, e.g., the Internet of Things. Alternative approaches are based on information-theoretic or physical-layer security, which are not affected by the computational capability of attackers. In \ac{PLA}, transmitters are differentiated only based on the electromagnetic characteristics of their transmission channels. 

\ac{PLA} has been studied in the literature for quite some time, using various features of received signals, such as \ac{CFR} and \ac{CIR}, to distinguish a legitimate user from a potential attacker, \cite{11003929}.
Recently, the \ac{AoA} of the signal has been shown to be a robust feature for \ac{PLA}, \cite{Pham2023,Srinivasan2024}.
In addition, user classification has been done using both classical statistical approaches and modern tools based on machine learning.

In parallel, wireless communications have evolved through the introduction of \acp{RIS} that, with their ability to shape the propagation environment, improve energy efficiency, reduce hardware complexity, and improve coverage. \acp{RIS} have also been considered to improve \ac{PLA}. Variable and random configurations can be set on the \ac{RIS} to generate challenge-response pairs and propose a challenge-response \ac{PLA} protocol based on the \ac{CSI}, \cite{Tomasin2022, tomasin2024analysis, globecom2023, icc1}. 
In \cite{Selim2023}, the authors consider \ac{CFR}-based \ac{PLA} in the presence of a hybrid \ac{RIS}, also capable of acting as a receiver and estimating the channels of impinging signals; thus, this estimate is exploited for authentication. Authentication in a scenario with an \ac{RIS} is studied also in \cite{Zhang2023}, however, also exploiting pre-shared keys used for asymmetric cryptography; thus, it cannot be considered working purely at the physical layer.
In \cite{Liu2024} \ac{PLA} based on the \ac{CIR} in a dynamic wireless communication environment, is studied, and convolutional neural networks are used to perform classification: this overcomes the limitations of the classical statistical approach based on hypothesis testing when the wireless channel is time-varying.

In this paper, we consider that signals received by Bob are reflected through a \ac{RIS} that he controls, and the legitimate transmitter, Alice, is equipped with a single antenna. In contrast, the adversary, Trudy, possesses multiple antennas and employs precoding techniques to attempt to bypass the verification process. Assuming Trudy has full knowledge of all channel matrices, we first determine her optimal attack strategy. We then examine the conditions under which Bob's channel estimation is identical regardless of whether Alice or Trudy is transmitting. When Trudy is limited to a single antenna, we derive conditions based on the angle of arrival at the RIS. Our analysis shows that under multipath propagation conditions to the RIS, the indistinguishability requirement cannot be satisfied. However, in the case of a single-path line-of-sight (LOS) scenario, Trudy can successfully impersonate Alice by transmitting from a different location. These findings are supported by numerical simulations. We also evaluate the system's security in situations where indistinguishability cannot be achieved.

The rest of the paper is organized as follows. Section~\ref{sec:mod} presents the system model. Section~\ref{sec:PLA} describes the \ac{PLA} mechanism and, then, in Section~\ref{sec:secan}, a security analysis is performed, focusing on conditions that make the attack indistinguishable from a legitimate signal. Numerical results are discussed in Section~\ref{sec:res} and, finally, conclusions are drawn in Section~\ref{sec:concs}.
\glsresetall
\section{System Model}
\label{sec:mod}

\begin{figure}[t!]
\centering
\includegraphics[width=90mm]{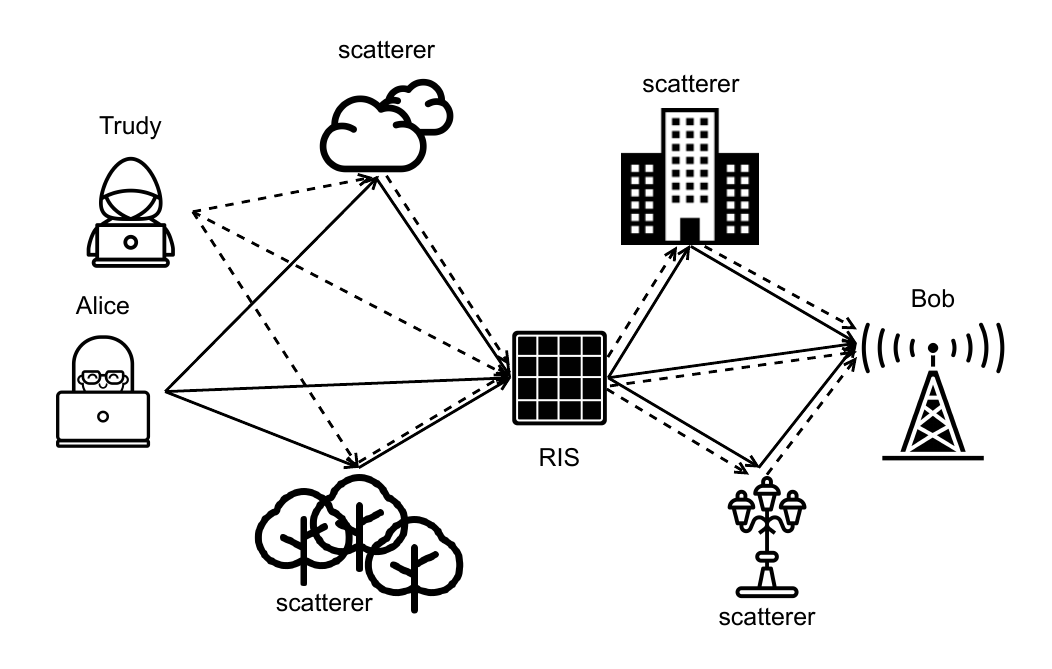}
\caption{System model.}
\label{fig:sys_mod}
\end{figure}

We consider the uplink scenario shown in Fig. \ref{fig:sys_mod}, where the \ac{BS} (Bob) aims to authenticate a \ac{UE} (Alice) in a \ac{simo} communication system, with Alice equipped with a single antenna and Bob with a \ac{ULA} of $M$ antennas. The signal transmitted by Alice reaches Bob through a \ac{RIS}, while a blockage obstructs the Alice-Bob direct link. An attacker device, Trudy, attempts to impersonate Alice by transmitting messages that Bob may mistake as originating from Alice. Trudy is equipped with a \ac{ULA} of $N_{\rm T}$ antennas. We also assume that no direct communication is possible between Trudy and Bob, and that all of her messages are transmitted through the RIS.

Transmissions occur at \ac{mmwave} frequencies. \ac{ULA} antennas are uniformly spaced by a distance $d = {\lambda}_c/2$, where $ {\lambda}_c$ is the carrier wavelength. Moreover, we assume that the field of view of Bob is $120^{\circ}$.

The \ac{RIS}, controlled by Bob, has $N$ reflecting elements spaced by the same distance $d$. The $n$-th element, $n=0,1,\hdots, N-1$, of the \ac{RIS} introduces a phase shift $\omega_n = e^{j\varphi_n}$ on the equivalent baseband signal and has unitary gain. The \ac{RIS} configuration matrix is defined as 
\begin{equation}
\bm{\Omega} = {\rm diag}\{[ e^{j\varphi_0},\ldots , e^{j\varphi_{N-1}}]\}.
\end{equation}

We denote the baseband equivalent vector for the channel from Alice to the \ac{RIS} as $\bm{f}\in\mathbb{C}^{N\times 1}$, the channel matrix from the \ac{RIS} to Bob as $\bm{G}\in\mathbb{C}^{M \times N}$.
Thus, the resulting Alice-\ac{RIS}-Bob cascaded channel is 
\begin{equation}\label{eq:cascadedCh}
    \bm{h}_{\mathrm{ARB}}=\bm{G\Omega f}\,.
\end{equation}
Alice transmits suitable pilot symbols to let Bob estimate the channel, which is used for authentication. The pilot signal is assumed to be known to Trudy. 

We denote as $\bm{T}$ the matrix of the channel from Trudy to the \ac{RIS}. To impersonate Alice, Trudy precodes the transmitted signal (including pilots) with vector $\bm{q}$ and the resulting Trudy-\ac{RIS}-Bob channel is then \begin{equation} \label{eq:HTR} \bm{h}_{\rm TRB} = \bm{G \Omega Tq} \in \mathbb{C}^{M \times 1}. \end{equation}
All channels ($\bm{f}$, $\bm{G}$, and $\bm{T}$) are time-invariant.

\subsection{Channel Model}

In the presence of objects around the transmitter and the receiver, the transmitted signal reaches the receiver through multiple paths. At the mmWave band, channels typically have only a few relevant paths; thus, we use a geometric model for their description. We define the $K$-size array response column vector for \ac{AoA} $\theta$ as
\begin{equation}
\label{eq:2}
 \bm{e}_K(\theta) = \frac{1}{\sqrt{K}} [1 , e^{-j \frac{2\pi}{{\lambda}_c} d \sin\theta}, \ldots,  e^{-j (K - 1) \frac{2\pi}{{\lambda}_c} d \sin\theta}]^T. 
 \end{equation}
For a generic channel with $L$ paths, we define the $L$-paths array response matrix with \ac{AoA} angles $\bm{\theta} = [\theta_1,...,\theta_L]^T$ as 
\begin{equation}
\bm{E}_N(\bm{\theta}) = [\bm{e}_N(\theta_1) , ..., \bm{e}_N(\theta_L)].
\end{equation}

Let $L_f$ be the number of paths between Alice and the \ac{RIS}, and $\phi_{f,l}$, $\theta_{f,l}$, and $\gamma_{f,l}$ represent the \ac{AoD} at Alice, the \ac{AoA} at the \ac{RIS}, and the complex path gain for the $l$-th path i.e., $l = 1,...,L_f$, respectively. Let us also define $\bm{\phi}_f = [{\phi}_{f,1},...,{\phi}_{f, L_{f}}]^T$ and $\bm{\theta}_f = [{\theta}_{f,1},...,{\theta}_{f, L_{f}}]^T$. Moreover, $\bm{1}_{L_f}$,  
$\bm{E}_N(\bm{\theta}_f)$, and $\bm{\Gamma}_f = \text{diag}([\gamma_{f,1}, ..., \gamma_{f,L_{f}}]^T)$ denote the $L$-size column vector of ones corresponding to Alice's array response matrix, the \ac{RIS} array response matrix, and diagonal path gain matrix, respectively. The baseband channel matrix between Alice and the \ac{RIS} is modeled as~\cite{masood2023inductive}
\begin{equation}
 \begin{split}
 \label{eq:3}
 \bm{f} = \sqrt{\frac{KN}{L_{f}}} \sum_{l=1}^{L_{f}}{\gamma_{f,l} \bm{e}_N(\theta_{f,l})\bm{e}_{1}^H(\phi_{f,l})}
      = \bm{E}_N(\bm{\theta}_f)\bm{\Gamma}_f \bm{1}_{L_f}\,.
\end{split}
\end{equation}

The \ac{RIS}-Bob channel matrix is modeled as
\begin{equation}
 \begin{split}
 \label{eq:G}
 \bm{G} = \bm{E}_M(\bm{\theta}_G)\bm{\Gamma}_G\bm{E}_N^H(\bm{\phi}_G)  \in \mathbb{C}^{M \times N}\,,
 \end{split}
 \end{equation} 
 where $\bm{\theta}_G$ and $\bm{\phi}_G$ are the vectors of \acp{AoA} to Bob and \acp{AoD} from the \ac{RIS}, and $\bm{\Gamma}_G$ is the diagonal matrix of path gains.
 
Similarly, the Trudy–\ac{RIS} channel is modeled as 
\begin{equation}
\begin{split}
\label{eq:6}
\bm{T} = \bm{E}_N(\bm{\theta}_t)\bm{\Gamma}_t\bm{E}_{N_t}^H(\bm{\phi}_t)  \in \mathbb{C}^{N \times N_{\rm T}},
\end{split}
\end{equation}
where $\bm{\theta}_t$ and $\bm{\phi}_t$ are the vectors of \acp{AoA} to the \ac{RIS} and \acp{AoD} from Trudy, and $\bm{\Gamma}_t$ is the diagonal $L_t \times L_t$ matrix of the $L_t$ path gains.
 
\subsection{Assumptions on Trudy}

Trudy is assumed to perfectly know all the channels, including the Alice-\ac{RIS} and \ac{RIS}-Bob channel matrices $\bm{f}$ and $\bm{G}$. This assumption is very generous to Trudy, because she typically is neither co-located with Alice nor Bob.
Moreover, the channels corresponding to $\bm{f}$ and $\bm{G}$ are only experienced in cascade through the \ac{RIS}. Note that Alice and Bob can easily estimate the overall cascaded Alice-\ac{RIS}-Bob channel, while it is harder for them, and even more so for Trudy, to estimate the individual channels represented by $\bm{f}$ and $\bm{G}$. Consequently, considering the attacker with complete channel knowledge will result in a conservative estimate of the security performance, corresponding to a worst-case condition for the legitimate receiver. 

We also assume that Trudy chooses the transmit power without restrictions.
Finally, we assume that neither Alice nor Bob knows the instantaneous channels with Trudy nor their statistics. In particular, Alice and Bob do not know where Trudy is located, so they cannot infer anything about the propagation of signals transmitted or received by Trudy.

\subsection{Communication-Optimal RIS Configuration}
\label{commetric}

Since the \ac{RIS} is used for communication purposes between Alice and Bob, its configuration should be optimized accordingly by Bob. We indicate the {\em communication-optimal \ac{RIS} configuration} maximizing the spectral efficiency as
\begin{equation}\label{commoptconf}
 \overlinebold{\Omega} = {\rm diag} (e^{j\bar{\varphi}_1}, \ldots, e^{j\bar{\varphi}_N}), 
\end{equation}
where $\bar{\varphi}_n$, $n=0, \ldots, N-1$, represent the communication-optimal phase shifts of the $N$ \ac{RIS} elements. 
Various works in the literature have proposed methods for optimizing the \ac{RIS} configuration. Here we consider the technique of \cite{Sayeed2023}.

\section{Physical Layer Authentication Mechanism}
\label{sec:PLA}

We consider a \ac{PLA} mechanism, where Bob aims at deciding between the two hypotheses 
\begin{align}
\mathcal{H}_0 &: \text{the signal comes from Alice,} \nonumber\\
\mathcal{H}_1 &: \text{the signal comes from the attacker Trudy.}\nonumber
\end{align}
To this end, the channel vector estimated by Bob operates as a distinguishing feature between the transmissions done by Alice and Trudy. 

The \ac{PLA} mechanism includes two phases, namely the association and verification phases. Since we assume that Bob does not know the cascade channel when Trudy is transmitting, we will not exploit this information for \ac{PLA}.

In the association phase, Alice transmits some known pilot signal $s_0$ to Bob, who exploits its knowledge to obtain a noisy estimate of $\bm{h}_{\rm ARB}$ that we denote $\bar{\bm{h}}$. We assume that such a phase is authenticated at a higher layer; thus, it provides a reliable  estimate of the Alice-Bob channel. The association phase has to be repeated every time the Alice-Bob channel changes.
In the subsequent verification phase, upon reception of a signal Bob estimates the channel over which such a signal traveled, assuming that $s_0$ was transmitted, and obtaining the estimate $\hat{\bm{h}}$.
Then, Bob performs a test on the obtained estimate to decide whether the transmitter was Alice or not.

Let $\bm{r}$ denote the signal received by Bob when Alice is transmitting. Assuming that Bob knows $s_0$ and the communication-optimal \ac{RIS} configuration $\overlinebold{\Omega}$, the received signal is $\bm{r} = \bm{h}_{\rm ARB} s_0 + \bm{n}$, 
where $\bm{n}$ is a circularly-symmetric complex Gaussian vector with zero mean and variance $\sigma_n^2$ per entry. Bob obtains an estimate of the channel as
\begin{equation}\label{estch}
\hat{\bm{h}} = \frac{\hat{\bm{r}}}{s_0}= \bm{h}_{\rm ARB} + \frac{\bm{n}}{s_0}.
\end{equation} 

Since we do not exploit any information on Trudy's channel for this test, we resort to the \ac{LT} on $\hat{\bm{h}}$, 
based on the norm-2 distance between the current channel estimate and that obtained in the association phase \cite{Baracca2012}, i.e.,
\begin{equation}\label{eq:mse}
\zeta = \|\hat{\bm{h}} - \bar{\bm{h}}\|^2.
\end{equation}
The \ac{LT} providing a decision $\hat{\mathcal H}$ between the two hypotheses is obtained by thresholding $\zeta$ as follows
\begin{subequations}
\label{testMSE}
\begin{equation}
\zeta < \tau: \; \hat{\mathcal H} = {\mathcal H}_0, \quad 
\zeta \geq \tau: \; \hat{\mathcal H} = {\mathcal H}_1,
\end{equation}  
\end{subequations}
where $\tau$ is a suitably chosen threshold.

\subsection{Security Metrics} 
\label{sec:secPer}
Two possible error events might occur in the authentication mechanism: the \ac{FA}, when Bob discards a message as forged by Trudy while it is coming from Alice, and the \ac{MD}, when Bob accepts a message coming from Trudy as legitimate. 

Specifically, an \ac{FA} occurs when, under hypothesis $\mathcal{H}_0$, $\zeta \geq \tau$, whereas, an \ac{MD} occurs when, under hypothesis $\mathcal{H}_1$, $\zeta < \tau$. As security metrics, we then consider the probabilities of \ac{FA} and \ac{MD}, i.e.
\begin{equation}
    P_{\mathrm{FA}} = \mathbb{P}[\zeta \geq \tau | \mathcal{H}_0] \,, \quad 
    P_{\mathrm{MD}} = \mathbb{P}[\zeta < \tau | \mathcal{H}_1] \,.\label{eq:pmd}
\end{equation}

\section{Security Analysis}
\label{sec:secan}

We now analyze the security of \ac{PLA} for the considered scenario. The obtained results will highlight how the structure of the channel, due to the few reflection paths, has an impact on the error probabilities of \ac{PLA}. First, we compute the optimal precoding vector for Trudy that maximizes the probability of her attack succeeding, i.e., maximizes the \ac{MD} probability. Then, we discuss the impact of the number of paths on the security.

Let us define the cascade channels when Alice and Trudy are transmitting as
\begin{equation}\label{aT}
\bm{c}_{\rm A} = \bm{E}_M(\bm{\theta}_G)\bm{\Gamma}_G\bm{E}_N^H(\bm{\phi}_G) \overlinebold{\Omega} \bm{E}_N(\bm{\theta}_f)\bm{\Gamma}_f\bm{1}_{L_f},
\end{equation}
\begin{equation}\label{a}
\begin{split}
    \bm{c}_{\rm T}  &= \bm{E}_M(\bm{\theta}_G)\bm{\Gamma}_G\bm{E}_N^H(\bm{\phi}_G) \overlinebold{\Omega} \bm{E}_N(\bm{\theta}_t)\bm{\Gamma}_t\bm{E}_{N_t}^H(\bm{\phi}_t)\bm{q} \\& = \bm{c}'_{\rm T}\bm{q}\,,
    \end{split}
\end{equation}
where $\bm{q}$ is the precoding vector used by Trudy to try to falsify Alice's channel. 
Then, the channel estimated by Bob when Alice is transmitting can be written as $\hat{\bm{h}}_A = \bm{c}_{\rm A}  + \bm{n}$, while the estimated channel when Trudy is transmitting with precoding vector $\bm{q}$ is $
\bm{\hat{h}}_T = \bm{c}'_{\rm T}\bm{q} + \hat{\bm{n}}$.

\subsection{Trudy Optimal Transmit Power}

Trudy's goal is to maximize the probability that Bob accepts her message as legitimate, i.e., to maximize $P_{\rm MD}$. Considering the likelihood \eqref{eq:mse} used in the \ac{LT}, Trudy must choose $\bm{q}$ to minimize $\zeta$, as Trudy knows the Alice-Bob cascade channel $\bm{c}_{\rm A}$. However, she does not know the noise of the estimate obtained by Bob in the association phase. Therefore, we obtain the following impersonation optimization problem 
\begin{equation}\label{minprob}
    \bm{q}^\star = \argmin_{\bm{q}}  \| \bm{c}'_{\rm T}\bm{q} - \bm{c}_{\rm A}\|^2 \,.
\end{equation}
Now, we have 
\begin{equation}
\begin{split}
\zeta &= ||\bm{c}'_{\rm T}\bm{q}-\bm{c}_{\rm A}||^{2} \\
&=  \bm{r}^{H}\bm{c}_{\rm A}-\bm{c}_{\rm A}^{H}\bm{c}'_{\rm T}\bm{q}-\bm{q}^H\bm{c}_{\rm T}^{'H}\bm{r}+\bm{q}^H\bm{c}_{\rm T}^{'H}\bm{c}'_{\rm T}\bm{q},
\end{split}
\label{eq:norm2}
\end{equation}
and by nulling the derivative with respect to $q$, the solution of the minimization problem \eqref{minprob} is 
\begin{equation}\label{solgen}
\bm{q}^\star =  \bm{c}^{'H}_{\rm T}(\bm{c}'_{\rm T}\bm{c}^{'H}_{\rm T})^{-1}   \bm{c}_{\rm A}.
\end{equation}

\subsection{Indistinguishability Conditions}

When $\zeta=0$, the Alice-Bob channel is indistinguishable from the Trudy-Bob channel, and Bob cannot detect an attack. 
Let us investigate which are the conditions under which this may occur.
Clearly, when Trudy is in the same position as Alice, they have the same channel to Bob. The interesting point here is to understand if there are other positions of Trudy that (together with some optimum precoding vector $\bm{q}$) provide the same indistinguishability condition. Such positions may exist, since Bob estimates only the {\em cascade channel} from Alice, and signals transmitted by Trudy pass through the same \ac{RIS} used by Alice. 
From \eqref{minprob} we note that indistinguishability is achieved when the system of complex linear equations
\begin{equation}
\bm{c}'_{\rm T} \bm{q} = \bm{c}_{\rm A}
\end{equation}
is solvable. However, determining general conditions on the Trudy-\ac{RIS} channel that ensure the solution is challenging. Therefore, in the following, we focus on the special case in which also Trudy has a single transmit antenna, for which a theoretical analysis is feasible. 

\subsection{Indistinguishability Conditions for $N_{\rm T}=1$}
 \label{subsec:indCon} 

Let us focus on the case in which Trudy has a single antenna and both Alice-\ac{RIS} and Trudy-\ac{RIS} channels have $L$ paths. Thus \eqref{a} becomes
\begin{equation}\label{a22}
    \bm{c}_{\rm T} = \bm{E}_M(\bm{\theta}_G)\bm{\Gamma}_G\bm{E}_N^H(\bm{\phi}_G) \overlinebold{\Omega} \bm{E}_N(\bm{\theta}_t)\bm{\Gamma}_t\bm{1}_Lq\,,
\end{equation}
and the precoding vector boils down to the scalar $q$.

To understand the conditions for indistinguishability in this case, let us define  $\bm{W} = \bm{E}_M^H(\bm{\theta}_G) \bm{E}_M(\bm{\theta}_G) \in \mathbb{C}^{L_G \times L_G}$ as the 
matrix with entry $[\bm{W}]_{ii}=M$ and
\begin{equation}
    [\bm{W}]_{ij} = 
    \sum_{m=1}^{M}{e^{-j(m-1)\kappa(\sin{\theta}_{G,i}-\sin{\theta}_{G,j})}}\,, \quad \mbox{for  $i\neq j$} 
    \label{eq:W}
\end{equation}
$\bm{z}_A$ as a $L_G$-size vector with entry $l_1 = 1, \ldots, L_G$
\begin{equation}\label{eq:zA}
[\bm{z}_A]_{l_1} = \sum_{l_2=1}^{L_f} \gamma_{f,l_2} \sum_{n=1}^{N} e^{-j[\kappa(n-1)\mu_{A,l_1 l_2} + \bar{\varphi}_{n}]},
\end{equation}
for  $\mu_{A,l_1 l_2} = (\sin \phi_{G,l_1} - \sin \theta_{f, l_2})$, and $\bm{z}_T$ as a $L_G$-size vector with entry 
\begin{equation}\label{eq:zT}
[\bm{z}_T]_{l_1} = \sum_{l_2=1}^{L_t} \gamma_{t,l_2}\sum_{n=1}^{N} e^{-j[\kappa(n-1)\mu_{T,l_1 l_2} + \bar{\varphi}_{n}]},
\end{equation} 
for  $\mu_{T,l_1 l_2} = \sin \phi_{G,l_1} - \sin \theta_{t, l_2}$. 
We also have
\begin{equation}\label{eq:t1}
\bm{c}_{\rm A}^{H}\bm{c}_{\rm A}  =  \bm{z}_A^H \bm{\Gamma}_G^H \bm{W} \bm{\Gamma}_G \bm{z}_A,
\end{equation}
\begin{equation}\label{eq:t2}
\bm{c}_{\rm T}^{'H}\bm{c}'_{\rm T} =  \bm{z}_T^H \bm{\Gamma}_G^H \bm{W} \bm{\Gamma}_G\bm{z}_T,
\end{equation}
\begin{equation}\label{eq:t3}
    \bm{c}_{\rm A}^H \bm{c}'_{\rm T} = \bm{z}_A^H \bm{\Gamma}_G^H \bm{W} \bm{\Gamma}_G\bm{z}_T,
\end{equation}
\begin{equation}\label{eq:t4}
    \bm{c}_{\rm T}^{'H} \bm{c}_{\rm A} = \bm{z}_T^H \bm{\Gamma}_G^H \bm{W} \bm{\Gamma}_G\bm{z}_A = (\bm{c}_{\rm A}^H \bm{c}_{\rm T}' )^{H}.
\end{equation}

Now, substituting \eqref{eq:t1}, \eqref{eq:t2}, \eqref{eq:t3}, and \eqref{eq:t4} into \eqref{eq:norm2}, and for $\Tilde{\bm{W}} =\bm{\Gamma}_G^H \bm{W} \bm{\Gamma}_G$, we have
\begin{equation}\label{eq:zeta2}
    \zeta = \bm{z}_A^H\Tilde{\bm{W}}\bm{z}_A - q\bm{z}_A^H \Tilde{\bm{W}} \bm{z}_T - q^*\bm{z}_T^H\Tilde{\bm{W}}\bm{z}_A + qq^* \bm{z}_T^H \Tilde{\bm{W}}\bm{z}_T \,.\end{equation}

Defining $b{=}\bm{z}_A^H \Tilde{\bm{W}} \bm{z}_A$, $c{=}\bm{z}_A^H \Tilde{\bm{W}} \bm{z}_T$, and $d{=}\bm{z}_T^H \Tilde{\bm{W}} \bm{z}_T$, \eqref{eq:zeta2} becomes
\begin{equation}\label{eq:mse0}
     \zeta = d|q|^2 - cq - (cq)^* + b\,.  
\end{equation}

We are now ready to investigate the indistinguishability condition. 
Replacing $q=\beta e^{j\alpha}$ in \eqref{eq:mse0}, such condition can be written as 
\begin{equation}\label{eqSol}
    d\beta^2 - 2|c|\beta \cos(\alpha + \rho) + b = 0\,, 
\end{equation}
with $c=|c|e^{j\rho}$. We firstly note that (by definition) $\zeta \geq 0$ and it is minimized for $\alpha^\star = -\rho$. Substituting $\alpha^\star$ in \eqref{eqSol}, we have $d\beta^2 - 2|c|\beta + b = 0$, which has solutions only if $|c|^2 - bd \geq 0$, or, equivalently, if
\begin{equation} \label{eqDiscr}
    |\bm{z}_A^H \Tilde{\bm{W}} \bm{z}_T|^2 \geq (\bm{z}_A^H \Tilde{\bm{W}} \bm{z}_A)(\bm{z}_T^H \Tilde{\bm{W}} \bm{z}_T).
\end{equation}
However, by the Cauchy-Schwarz inequality
\begin{equation}
    |\bm{z}_A^H \Tilde{\bm{W}} \bm{z}_T|^2 \leq (\bm{z}_A^H \Tilde{\bm{W}} \bm{z}_A)(\bm{z}_T^H \Tilde{\bm{W}} \bm{z}_T),
\end{equation}
and thus \eqref{eqDiscr} must hold with equality. However, this happens if and only if  $\sqrt{\Tilde{\bm{W}}}\bm{z}_A$ and $\sqrt{\Tilde{\bm{W}}}\bm{z}_T$ are linearly dependent. Note that this does not generally imply $\bm{z}_A$ and $\bm{z}_T$ to be linearly dependent unless $\Tilde{\bm{W}}$ is a full rank matrix. By definition, the rank of $\Tilde{\bm{W}}$ is the same of $\bm{W}$ (due to $\bm{\Gamma}_G$ being diagonal), which is full rank if and only if the vectors $\{\bm{e}_M(\theta_{G,i})\}_{i=1}^{L_G}$ (i.e., the columns of $\bm{E}_M(\bm{\theta}_G)$) are linearly independent. This condition is satisfied when $L_G \leq M$ and the angles $\theta_{G,i}$ related to the different paths are distinct, i.e.,  $\sin\theta_{G,i} \neq \sin\theta_{G,j}$, $\forall i,j = 1, \ldots, L_G$, with $i \neq j$. Since each entry of $\bm{W}$ is given by the inner product of array response vectors \eqref{eq:W}, which depend only on $\sin(\cdot)$ and are periodic over $\pi$ for ULAs with half-wavelength spacing, we must have 
\begin{equation}
\theta_{G,i} \neq \theta_{G,j} + u\,\pi,
\end{equation}
for any integer $u$. Since we assume Bob has a field of view of $\frac{2}{3}\pi$, we are also ensuring $\bm{W}$ to be full rank when $L_G \leq M$. In this case, it can be stated that \eqref{eqDiscr} holds with equality if and only if $\bm{z}_A$ and $\bm{z}_T$ are linearly dependent.
From the definitions in \eqref{eq:zA} and \eqref{eq:zT}, we conclude that the indistinguishability conditions require that Alice and Trudy have the same number of paths ($L_t=L_f$), the \ac{AoA} angles at the \ac{RIS} corresponding to Alice and Trudy match exactly, yielding
\begin{equation}\label{zAzTlinDep1}
\sin \theta_{f,l} = \sin \theta_{t,l}, \quad l=1, \ldots, L_t = L_f\,,
\end{equation}
and their path gains are proportional, i.e.,
\begin{equation}\label{zAzTlinDep2}
\gamma_{f,l} = \lambda\, \gamma_{t,l}\,, \quad l=1, \ldots, L_t = L_f\,.
\end{equation}
These are then the indistinguishability conditions for $N_{\rm T}= 1$.

\subsection{Single-Path \ac{RIS}-Bob Channel}
\label{subsec:sp}

When the \ac{RIS}–Bob channel is single-path ($L_G{=}1$), $\bm{z}_A$ and $\bm{z}_T$ collapse to complex scalars. This dimensionality reduction significantly simplifies the attacker's task, as linear dependence now can be trivially achieved in $\mathbb{C}$, where any two non-zero scalars are always linearly dependent if one is a scaled version of the other. 

Hence, it becomes easier for the attacker to find values of $\alpha$ and $\beta$ such that \eqref{eqSol} is satisfied. Indeed, in this case, even when Trudy does not show the same angles and path gains of Alice ($z_T \neq z_A$), indistinguishability can still be achieved by appropriately tuning $\alpha$ and $\beta$ so that \eqref{eqSol} holds. In formulas, this happens for 
\begin{equation}
\alpha = -\rho + u\pi, u \mbox{ even}, \alpha \in [-\pi, \pi], \mbox{ and }
\beta = \frac{|z_A|}{|z_T|}
\end{equation}
or 
\begin{equation}
\alpha = -\rho + u\pi, u \mbox{ odd}, \alpha \in [-\pi, \pi], \mbox{ and } \beta = -\frac{|z_A|}{|z_T|}.
\end{equation}

The case $L_G{=}1$ inherently poses a higher impersonation risk, as it offers fewer spatial degrees of freedom to differentiate between Alice and Trudy. 

This result could also be directly inferred from the structure of the cascaded channels in \eqref{aT} and \eqref{a22}. Since the common term $\bm{E}_M(\bm{\theta}_G)\bm{\Gamma}_G\bm{E}_N^H(\bm{\phi}_G)$ of the \ac{RIS}-Bob channel has rank $1$, the cascaded channels lie in the same one-dimensional subspace. Therefore, no matter how different Trudy’s and Alice’s angles and path gains are, once they pass through it, the result is always confined to a single spatial direction, limiting Bob's ability to distinguish between them. In fact, any differences in Alice and Trudy transmissions are effectively collapsed into a single direction by the rank-one projection of $\bm{G}$ and, then, Trudy can more easily mimic Alice's cascaded channel. 

\section{Numerical Results}
\label{sec:res}
In this section, we assess the performance of the considered
authentication method investigating both single-path (i.e., $L_G = 1$) and multipath (i.e., $L_G = 3$) scenarios for the \ac{RIS}-Bob channel. We consider $L_f=L_t=3$ and path gains ${\gamma_{f,l}}$, ${\gamma_{G,l}}$, and  ${\gamma_{t,l}}$ distributed as ${\mathcal{CN}}(0,1)$.  We assume that the angles at the \ac{RIS} and the \acp{AoD} from the transmitters are uniformly distributed in $\left[ -\frac{\pi}{2}, \frac{\pi}{2} \right]$, while the \acp{AoA} at Bob are uniformly distributed in the range $\left[ -\frac{\pi}{6}, \frac{\pi}{6}\right]$. Angles and gains are generated independently for Alice and Trudy. Bob is equipped with $M \in \{4, 8, 16, 32\}$ antennas, while Alice and Trudy are single-antenna devices. The number of \ac{RIS} elements is $N=64$. 

\begin{figure}
\vspace{0.1cm}
    \centering    
    \includegraphics[width=\columnwidth]{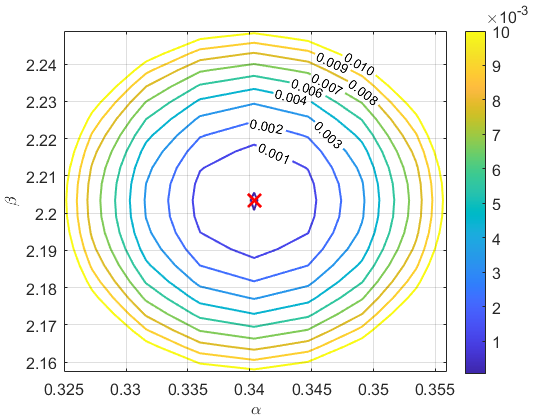}
    \caption{Contour plot of $\zeta$ (under hypothesis $\mathcal{H}_1$) for $L_G{=}1$, $L_f{=}L_t{=}3$, $M{=}16$, $N{=}64$. The red cross marks the values of $\alpha$ and $\beta$ that minimize $\zeta$. We consider different angles and path gains for the Trudy-\ac{RIS} and Alice-\ac{RIS} channels.} 
    \label{fig:sim_vs_anSP}
\end{figure}
Fig. \ref{fig:sim_vs_anSP} shows a contour plot of the test function $\zeta$ under attack conditions for a single-path \ac{RIS}-Bob channel  (i.e., $L_G=1$). Note that different angles and path gains for the Trudy-\ac{RIS} and Alice-\ac{RIS} channels are considered. The red cross marks the values of $\alpha$ and $\beta$ that minimize $\zeta$: when Trudy chooses the value of $q^\star$ corresponding to these optimal values of $\alpha$ and $\beta$, we have $\zeta = 0$. 

\begin{figure}
\vspace{0.1cm}
    \centering    
    \includegraphics[width=\columnwidth]{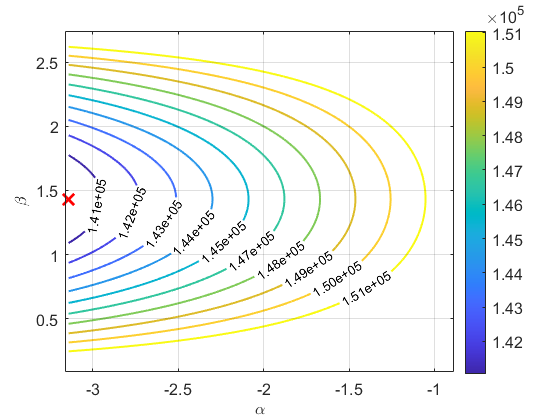}
    \caption{Contour plot of $\zeta$ (under hypothesis $\mathcal{H}_1$) for $L_G{=}L_f{=}L_t{=}3$, $M{=}16$, $N{=}64$. The red cross marks the values of $\alpha$ and $\beta$ that minimize $\zeta$. We consider different angles and path gains for the Trudy-\ac{RIS} and Alice-\ac{RIS} channels.} 
    \label{fig:sim_vs_anMP}
\end{figure}
Similarly, Fig. \ref{fig:sim_vs_anMP} shows a contour plot of the test function $\zeta$ under attack conditions for $L_G=3$. Comparing Figs. \ref{fig:sim_vs_anMP} and \ref{fig:sim_vs_anSP}, we observe that, for $L_G>1$, even if Trudy uses the optimal $q^\star$, the resulting minimum of the test function $\zeta$ is strictly greater than zero. This confirms that, unlike the scenario with $L_G=1$, perfect impersonation becomes impossible to achieve. Indeed, the presence of $L_G$ paths increases the rank of the \ac{RIS}–Bob channel matrix, thereby introducing additional spatial diversity that makes it harder for Trudy to align her cascade channel with that of Alice by setting the proper $q^\star$.

\begin{figure}
\vspace{0.1cm}
    \centering    
    \includegraphics[width=\columnwidth]{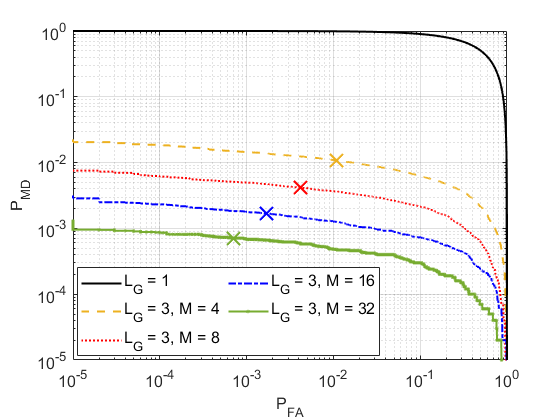}
    \caption{DET curves for different value of $M$, $L_G \in \{1, 3\}$. 
    The crosses mark the points for which $P_{\rm MD} = P_{\rm FA}$.} 
    \label{fig:DET}
\end{figure}

The result is also confirmed by Fig.~\ref{fig:DET}, which shows the detection error trade-off (DET) curves for different values of $M$ and $L_G \in \{1, 3\}$.
The crosses mark the points for which $P_{\rm MD} = P_{\rm FA}$. All the curves show that reducing $P_{\rm FA}$ results in an increase in $P_{\rm MD}$, and vice versa. It can also be noticed that for $L_G=1$, we have $P_{\rm MD} = 1-P_{\rm FA}$, regardless of the number of Bob's antennas $M$. In fact, in this case, Trudy can always find an attack strategy that yields to indistinguishability with Alice; thus the probability that Bob decides for hypothesis $\mathcal H_1$ (i.e., attack condition) is the same irrespective of who is transmitting. 
For $L_G>1$, instead, the optimal attack does not usually lead to indistinguishability (since the \acp{AoA} from Trudy and Alice are independent). Indeed, the DET curves do not start from the top-left corner as is typically the case. This is due to the statistical nature of the test and imperfections in Trudy’s impersonation of Alice. In fact, when $L_G>1$, the perfect alignment between Trudy's and Alice's cascaded channels is not achievable, even if Trudy uses $q^\star$. Hence, the minimum achievable $P_{\rm MD}$ is strictly less than $1$, emphasizing a significant limit on the success of the impersonation attack. Hence, we can conclude that a higher $L_G$ enhances authentication robustness by limiting the ability of Trudy to fully mimic Alice’s cascaded channel. Moreover, we observe that, as $M$ increases, the DET curves move towards smaller $P_{\rm MD}$ for a target $P_{\rm FA}$. This shows that having more receive antennas allows for better distinction between Alice and Trudy. 

\section{Conclusions}
\label{sec:concs}
We analyzed the security of a \ac{RIS}-assisted PLA scheme in scenarios with no direct link between the transmitter and the receiver, and multipath propagation conditions of the channels to and from the \ac{RIS}. Assuming the worst case scenario of an attacker Trudy having full channel knowledge, we determined her optimal attack strategy. Then, we examined the conditions under which Bob’s channel estimation may have the same statistics regardless of whether Alice or Trudy is transmitting, deriving the conditions based on the \acp{AoA} at the \ac{RIS} for single antenna attacker. Numerical results show that when the \ac{RIS}–Bob channel is single-path, impersonation is feasible even with mismatched channel parameters. Conversely, increasing the number of \ac{RIS}–Bob paths significantly enhances authentication robustness by limiting the attacker’s ability to mimic the legitimate user.

\bibliographystyle{IEEEtran}
\bibliography{biblio}

\end{document}